\documentclass[pdflatex,sn-mathphys-num]{sn-jnl}

\usepackage{graphicx}%
\usepackage{multirow}%
\usepackage{amsmath,amssymb,amsfonts}%
\usepackage{amsthm}%
\usepackage{mathrsfs}%
\usepackage[title]{appendix}%
\usepackage{xcolor}%
\usepackage{textcomp}%
\usepackage{manyfoot}%
\usepackage{booktabs}%
\usepackage{algorithm}%
\usepackage{algorithmicx}%
\usepackage{algpseudocode}%
\usepackage{listings}%
\usepackage{changes}
\definechangesauthor[name={Xie zhongyi}, color=red]{Xie}

\theoremstyle{thmstyleone}%
\theoremstyle{thmstyletwo}%

\theoremstyle{thmstylethree}%

\begin{document}

\title[Article Title]{Evolution of Correlated Electrons in ${\rm La_3Ni_2O_7}$ at Ambient Pressure: a Study of Double-Counting Effect}


\author[1]{Zhong-Yi Xie}\email{xiezhy9@mail2.sysu.edu.cn}

\author[1]{Zhihui Luo}\email{luozhh59@mail.sysu.edu.cn}

\author[1]{W\'{e}i W\'{u}}\email{wuwei69@mail.sysu.edu.cn}

\author*[1]{and Dao-Xin Yao}\email{yaodaox@mail.sysu.edu.cn}

\affil[1]{Guangdong Provincial Key Laboratory of Magnetoelectric Physics and Devices, State Key Laboratory of Optoelectronic Materials and Technologies, Center for Neutron Science and Technology, School of Physics, Sun Yat-sen University, Guangzhou, Guangdong 510275, China}


\abstract{We employ cluster extension of dynamical mean-field theory (CDMFT) to systematically investigate the impact of double counting corrections on the correlated electronic structure of ${\rm La_3Ni_2O_7}$ under ambient pressure. By adjusting double-counting parameters, while maintaining a fixed Fermi surface, we observe a pronounced orbital-selective density of states change: the $d_{z^2}$ orbital undergoes significant variation near the Fermi level with increasing $E_{dc}^z$, while the $d_{x^2-y^2}$ orbital remains essentially unchanged throughout the entire range. Analysis of renormalization factor show the monotonic dependence with double counting in both $d_{z^2}$ and $d_{x^2-y^2}$ orbital, and it also identifies an optimal double counting window in $d_{z^2}$ orbital aligns with experimental values. We also find the interlayer Matsubara self energy exhibits non-monotonic dependence on $E_{dc}^z$, deviating from theoretical predictions. This anomaly is attributed to the metallization of oxygen-bridged pathways, which disrupts the prerequisite for charge transfer via apical oxygen. Our results establish $E_{dc}$ as a critical control parameter for correlated electronic structure in ${\rm La_3Ni_2O_7}$ and provide a computational framework for resolving orbital-dependent correlation effects in layered materials.}




\maketitle

\section{Introduction}

The dynamical mean-field theory (DMFT) has emerged as an established methodology for calculating the electronic structure of strongly correlated materials, with recent advancements demonstrating its effectiveness in resolving complex many-body interactions that traditional density functional theory (DFT) approximations fail to capture\cite{dft+dmft_1,dft+dmft_2,dft+dmft_3}.  
However, an underlying problem of this approach comes from the estimation of the double-counting energy ($E_{dc}$), which is designated to remove the equivalent contribution 
within the correlated subspace that is already accounted for at the DFT level. 
This issue appears when the system contains both correlated and uncorrelated electrons. A common example would be the transition metal compounds, where the charge transfer energy ($E_{CT}$) between the correlated ions and its ligands becomes the charateristic energy scale\cite{ect}.
Here the choice of $E_{dc}$ will directly affect $E_{CT}$, which is even pronounced in the charge-transfer limit $E_{CT}\rightarrow 0$.
Several double-counting choices have been proposed \cite{edc_1,edc_2,edc_3} but an exact prescription is unattainable for the general cases.

Recently, transition temperature (T$_c$) reaching $\sim$80 K under high pressure was found in the bilayer nickelate ${\rm La_3Ni_2O_7}$ \cite{Nisc}. The average valence state of nickel (Ni) atoms in ${\rm La_3Ni_2O_7}$ is ${\rm Ni^{2.5+}}$ ($3d^{7.5}$), which contains the fully-filled $t_{2g}$ and partial filling of $e_g$ orbitals. The fundamental building block of ${\rm La_3Ni_2O_7}$ consists of two quasi-two-dimensional NiO$_2$ planes, which are coupled via $\sigma$-bonding of $d_{z^2}$ mediated by apical O-2$p_z$ orbital.  The half-filling of $d_{z^2}$ orbital suggests an effective interlayer superexchange coupling that is believed to be crucial for the superconductivity\cite{spex}. The superexchange coupling is  regulated by the hopping  $d_{z^2}$-$p$ \cite{edp} with the cost of the energy difference $\Delta_{dp}$.
Therefore, a careful choice of the $E_{dc}$ would be important to correctly account for the superconducting propertities.

In this paper, we systematically investigate the impact of double counting energy on the correlated electronic structure for ${\rm La_3Ni_2O_7}$  under ambient pressure. A charge-transfer 11 orbital model of ${\rm La_3Ni_2O_7}$ that contains 4 Ni-d oribtals (2 Ni atoms with two orbitals $d_{x^2-y^2}$, $d_{z^2}$  per Ni) and 7 ligand oxygen p orbitals is used, which is solved self-consistently under CDMFT. To ensure physical validity, besides explicitly adjucting the $E_{dc}$ values in a plausible range, we also  rigorously compare the calculated Fermi surface (FS) profile to match that from the experimentally angle-resolved photoemission spectroscopy (ARPES)\cite{ARPES}. With this, we investigate the evolution of electron density, density of states (DOS) and non-local interlayer correlation to gain a comprehensive understanding of the correlation feature. Also, a comparison to two commonly used double-counting schemes, the fully localized limit (FLL)\cite{edc_3} and Held formulas\cite{AMF}, are presented.

\section{Model and Method}

We consider an 11-band Hubbard model that can be written as:

\begin{equation}
	H = H_0 +H_U - \sum_{i,\alpha ,\sigma}E_{dc}^{\alpha}
	n_{i\alpha \sigma}^{d}
\end{equation}
\begin{align}
	H_0 = \sum_{i,j,\alpha,\beta,\sigma} t_{i,j,\alpha,\beta} \, d_{i\alpha\sigma}^{\dagger} \, p_{j\beta\sigma} + 
	\sum_{i,j,\alpha,\beta,\sigma} t_{i,j,\alpha,\beta} \, p_{i\alpha\sigma}^{\dagger} \, p_{j\beta\sigma} - \sum_{i,\alpha,\sigma} \mu_{\alpha} \, n_{i\alpha\sigma}
	\label{eq:hamiltonian}
\end{align}
\begin{align} 
	\nonumber
	H_U = &U\sum_{i,\alpha} n_{i\alpha\uparrow}^d n_{i\alpha\downarrow}^d 
	+ U^{\prime}\sum_{i,\alpha<\beta,\sigma} n_{i\alpha\sigma}^d n_{i\beta\bar{\sigma}}^d +\left(U^{\prime}-J_H\right)\sum_{i,\alpha<\beta,\sigma} n_{i\alpha\sigma}^d n_{i\beta\sigma}^d +
	\\&J_H\sum_{i,\alpha \neq \beta } (d_{i\alpha\uparrow}^{\dagger} d_{i\alpha\downarrow}^{\dagger} d_{i\beta\uparrow} d_{i\beta\downarrow}- 
	d_{i\alpha\uparrow}^{\dagger} d_{i\alpha\downarrow} d_{i\beta\downarrow}^{\dagger} d_{i\beta\uparrow})  
	\label{eq:hamiltonian_U}
\end{align}

Here $H_0$ is the tight-binding Hamiltonian determined out of our DFT calculation which implemented in the Vienna ab initio simulation package (VASP)\cite{24,vasp}. The projector augmented-wave (PAW) method\cite{paw} with a 600 eV plane-wave cutoff is adopted. The generalized gradient approximation (GGA) of Perdew-Burke-Ernzerhof (PBE)\cite{26} is used for exchange-correlation functional. $t_{i,j,\alpha ,\beta }$ denote hoppings between electrons on sites $(i,j)$ and orbital $(\alpha ,\beta )$ (can be either Ni-d or O-p orbitals). $d_{\alpha ,i,\sigma }^{\dagger }$($p_{\alpha ,i,\sigma }^{\dagger }$) is the creation operator for electrons on $\alpha \in 3d (\in 2p)$  orbital. $\mu $ is the chemical potential of $\alpha $-orbital. The schematic diagram of this model can be referred to in \cite{model}, and the basis is defined as: $\Phi = (d_{z_1},d_{z_2},d_{x_1},d_{x_2},p_{x_1},p_{x_2},p_{y_1},p_{y_2},p_z,p_{z_1}^{\prime},p_{z_2}^{\prime})^T$. 1,2 label the bilayer , $d_z$, $d_x$ label $d_{z^2}$, $d_{x^2-y^2}$ orbitals, $p_x$ and $p_y$ label the in-plane O's p-orbital, $p_z$ labels the inner apical O's p-orbital and $p_{z}^{\prime}$ labels outer apical O's p-orbital. $H_U$ is the Kanamori Coulomb interaction term, $U$ is the Hubbard interaction between two electrons on the same d-orbital ($d_{x^2-y^2}$ or $d_{z^2}$) and $U^{\prime}$ is for that on two different d-orbitals. $U^{\prime} = U-2J_H$ is adopted, where $J_H$ is the Hund's coupling, and intensity of spin flip term is same as pair hopping term. $E_{dc}$ is the double counting term to be subtracted in the DMFT. 

We solve the lattice model within the framework of DMFT. DMFT is a powerful non-perturbative approach that maps a lattice problem onto an effective Anderson single-impurity model, which is then solved self-consistently\cite{13,14,15}. This mapping becomes exact in the limit of infinite dimensions but it is also precise enough in two-dimensional system, and neglects all nonlocal contributions to the retarded self-energy. So the retarded self energy and Green's function is purely local, i.e., $\Sigma_{imp}\left(\omega\right)=\Sigma\left(\omega\right);G_{imp}\left(\omega\right)=G_{loc}\left(\omega\right)=\frac{1}{N_k}\sum_{k}{G(k,\omega)}$. Furthermore, we employed CDMFT to incorporate non-local interactions between layers\cite{16,17,18}, which captures the interlayer correlation properties in La$_3$Ni$_2$O$_7$ materials. Specifically, four orbitals ($d_{z_1},d_{z_2},d_{x_1},d_{x_2}$) are explicitly treated to form the cluster basis because of their predominance around the Fermi surface, while the electrons from remaining orbitals are encapsulated through a hybridization function that mediates their interactions with the impurity electrons.

In our CDMFT study, we use $U$ = 8 eV and $J_H$ =  1 eV to calculate an effective impurity model with the 2 $\times$ 2 = 4 orbitals which could introduce nonlocal interlayer coupling under $\beta = 25$. We use quantum Monte Carlo as the impurity solver, its number of steps is set to 10000 and length of step is set to 10000 which keeps the correction time $\sim $ 1s. Here, we use the open-source TRIQS\cite{27} package and its continuous-time quantum Monte Carlo\cite{28} for CDMFT and its impurity solver, respectively.

Given that there are two types of inequivalence correlated orbitals in the model, a consideration of the orbital-dependent double-counting energy for both $E_{dc}^x$, $E_{dc}^z$ would be more reasonable for the real situation but on the other hand bring an  expansion of the adjustable paramenter range.
To simplify the consideration, we careful align our DMFT Fermi surface profile with that from the ARPES for each tunning.
We find this procedure makes $E_{dc}^x$ and $E_{dc}^z$ depend on each other, but on the other hand, the site energy of apical $p_z$ orbital has to vary as well. That means for each tuning, the fitting of FS  can determine a set of parameters ( $E_{dc}^z$, $E_{dc}^x$, $\mu_{pz}$).
Therefore, in the following, we only present  the relevant physical quantities as a function of $E_{dc}^z$.

\section{Results}

\begin{figure}
	\centering
	\includegraphics[width=1\linewidth]{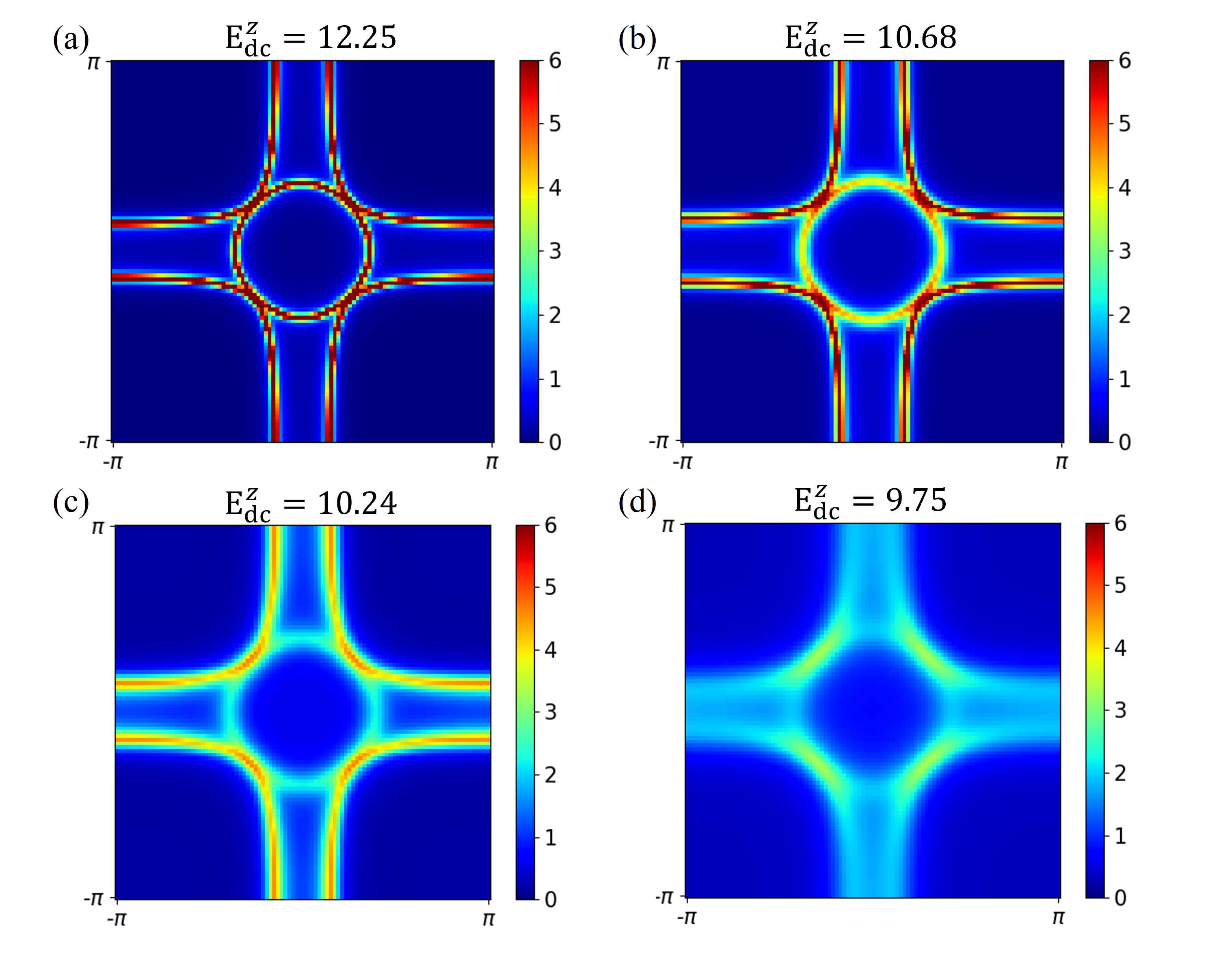}
	\caption{The evolution of Fermi surface with decreasing the double-counting energies $E_{dc}^z$. Here the Fermi surface profile is manually adjusted to match that from the ARPES result\cite{ARPES}, which leads to only one independent parameter among ($E_{dc}^z$, $E_{dc}^x$, $\mu_{pz}$), see main text for details.}
	\label{fig:fs}
\end{figure}

We first demonstrate the Fermi surface as a function of $E_{dc}^z$, as shown in Fig.~\ref{fig:fs}. One can see that only two pockets $\alpha,\beta$ show up, with each profile and position  in precise argreement with the ARPES result\cite{ARPES} under ambient pressure. The major difference comes from the coherence. When $E_{dc}^z$ is large, as shown in Fig.~\ref{fig:fs}(a), there is well-defined quasiparticle dispersion of both pockets, indicating Fermi liquid behavior. As decreasing $E_{dc}^z$, the incoherence is gradually enhanced, followed by the broadening of the spectrum, which is even pronounced in Fig.~\ref{fig:fs}(d).

\begin{figure}
	\centering
	\includegraphics[width=1\linewidth]{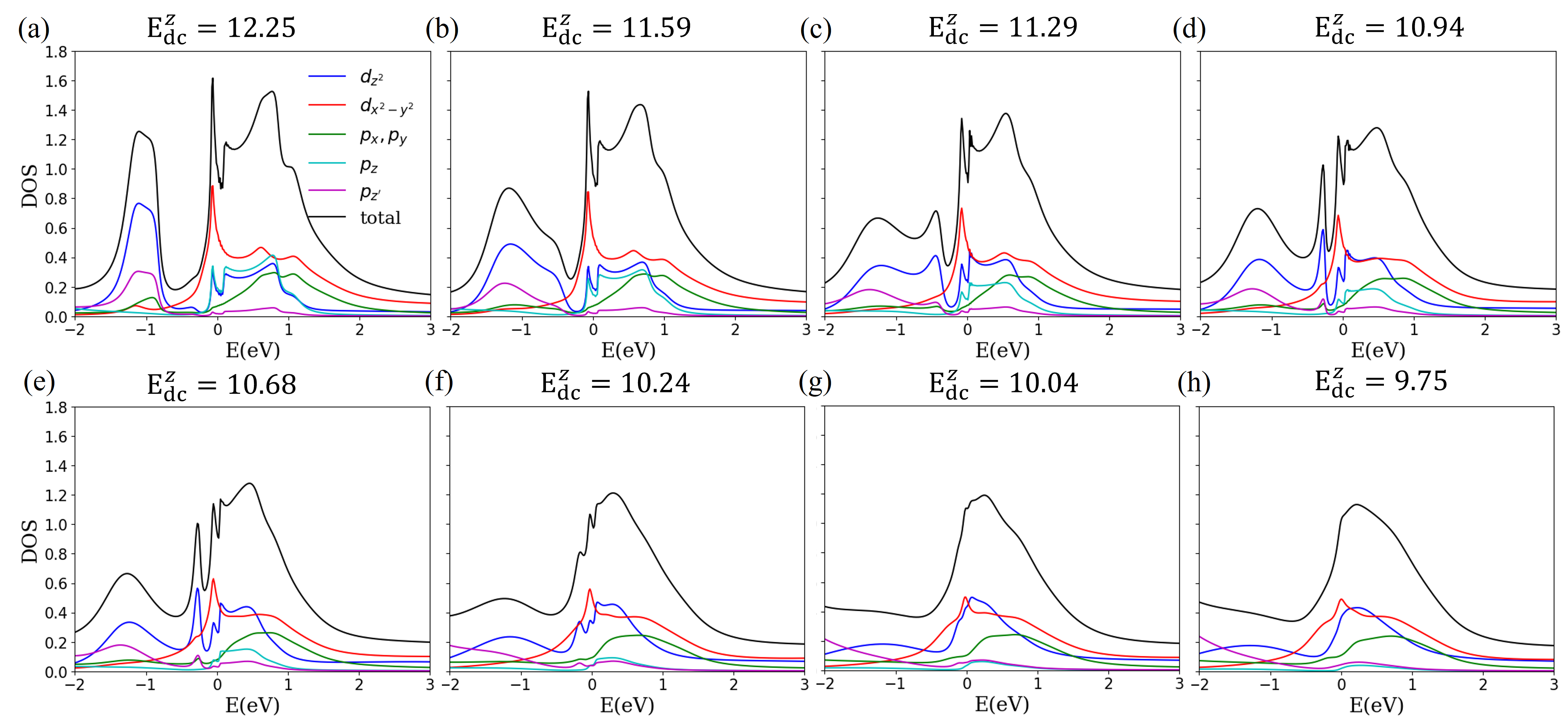}
	\caption{Evolution of DOS as a function of $E_{dc}^z$.}
	\label{fig:dos}
\end{figure}

To better understand the above features, we next present the evolution of density of states (DOS) around Fermi energy ($E_{\rm F}$) as a function of $E_{dc}^z$, as shown in Fig.~\ref{fig:dos}. In Fig.~\ref{fig:dos}(a), which corresponds to that in Fig.~\ref{fig:fs}(a), the DOS exhibits two major branches: one centers around -1\ eV and the other spans around 0-2\ eV, and they are separated by a gap of $\sim$0.8\ eV. Orbital-resolved DOS further reveals that the low energy branch is mostly associated with a mixing of $d_{z^2}$ orbital with apical O-$p_z$ orbital that outside the bilayer, while the other branch contains complex mixing of all orbitals. Moreover, a notable dip is observed precisely at $E_{\rm F}$, which is shaped by a sharp tip beneath $E_{\rm F}$ from $d_{x^2-y^2}$ orbital, and an above one from the antibonding band of $d_{z^2}$ orbital.
As decreasing $E_{dc}^z$, the low-energy branch is broadened and then shows a signal of splitting at $E_{dc}^z$=11.29 (Fig.~\ref{fig:dos}(c)). Further decreasing $E_{dc}^z$ leads the upper splitted band to merge into the high-energy branch. Eventually, all fine features are invisible with only one plain peak characterizes the spectrum, as illustrated in Fig.~\ref{fig:dos}(f-h). No doubt that $E_{dc}^z$ in this range is too far away from the real material.
However, even in the physical plausible range from Fig.~\ref{fig:dos}(a)-(e), there is drastic evolution of DOS  with  $E_{dc}^z$ only spans $\sim$ 1.5\ eV. This highlights  a sensitive dependence of the correlation feature on the double-counting energies.

The tip at $E_{\rm F}$ persists from Fig.~\ref{fig:dos}(a)-(e) indicates an intrinsic connection to the FS topology.
To gain insight, we  further inspect $A({\rm k},E)$ spectrum, which reveals that the sharp tip beneath $E_{\rm F}$ is mostly associated with the nearly flat band around $\Gamma$-X path (no shown), which can further track back to the proximity of $\beta$ pocket to the X ($\pi,0$) point, as shown in Fig.~\ref{fig:fs}.
It is interesting to note that, both ARPES and tunneling experiments\cite{31,32,33} reveal the presence of a gap at $E_{\rm F}$ that is in line with our calculation.

\begin{figure}
	\centering
	\includegraphics[width=1\linewidth]{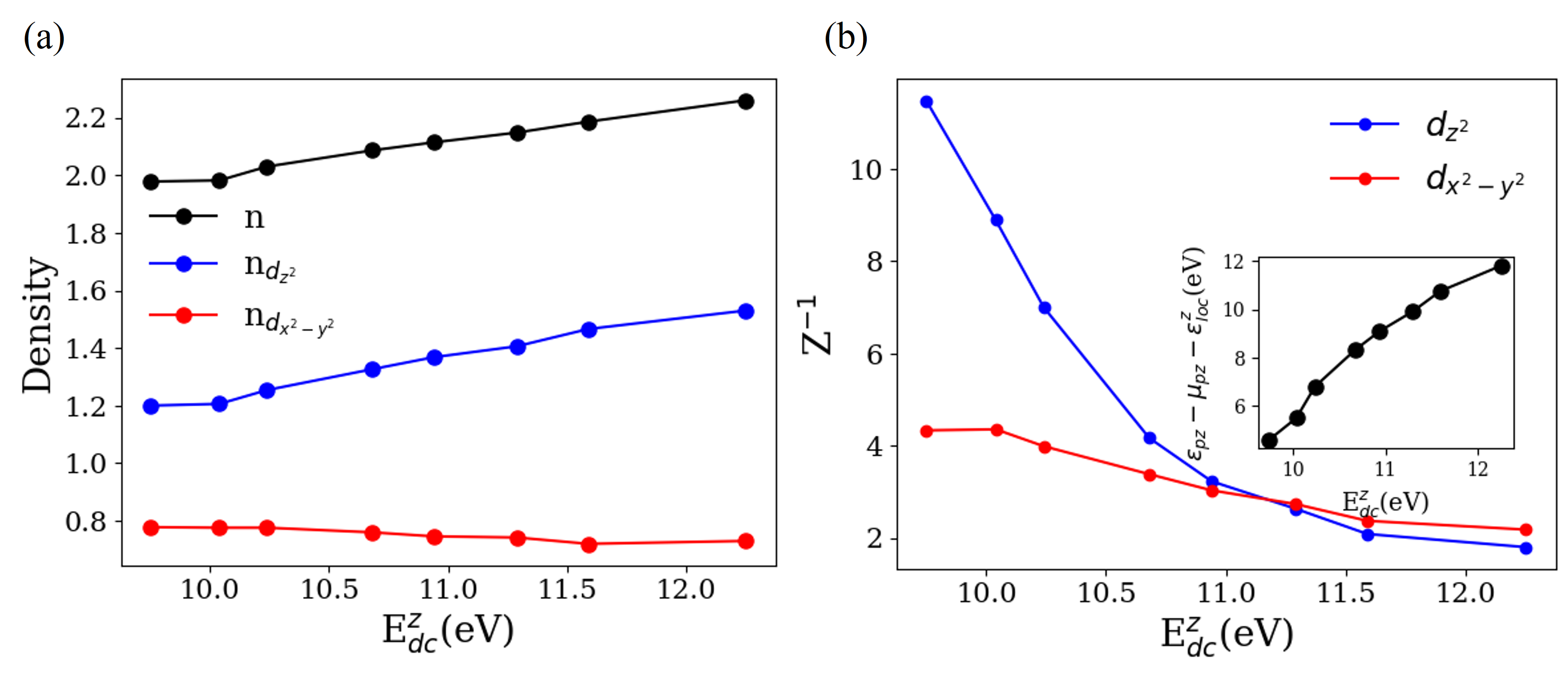}
	\caption{(a) Electron densities as a function of $E_{dc}^z$. (b) Renormalization factor $Z^{-1}$ for $d_{z^2}$ and $d_{x^2-y^2}$ orbitals  as a function of $E_{dc}^z$. Inset show the quantity $\epsilon_{pz}-\mu_{pz}-\epsilon_{loc}^z$ as a function of $E_{dc}^z$, which is proportional to the charge-transfer energy $E_{dp}$  and can be regarded as a reference.}
	\label{fig:density}
\end{figure}

Fig.~\ref{fig:density}(a) displays the electron density evolution of d-orbital states as a function of $E_{dc}^z$. The analysis reveals that the $d_{x^2-y^2}$ orbital exhibits negligible sensitivity to $E_{dc}^z$, whereas the $d_{z^2}$ orbital dominates the electronic density modulation. Specifically, the monotonic increase of $n_{d_{z^2}}$ with $E_{dc}^z$ originates from the growth of the $d_{z^2}$ bonding-state peak near -1 eV as $E_{dc}^z$ increase shown in Fig.~\ref{fig:dos}. In contrast, DOS of the $d_{x^2-y^2}$ orbital below the Fermi level remains largely unaffected by $E_{dc}^z$, indicating its weak role in this energy-dependent charge redistribution.

The correlation effect can be quantified using the quasiparticle mass renormalization factor $\frac{m^*}{m}=Z^{-1}=1 - \frac{\partial Im\Sigma (i\omega )}{\partial i\omega}\Big| _{i\omega \rightarrow 0}$ and we perform the renormalization factor varying with $E_{dc}^z$ in Fig.~\ref{fig:density}(b). As the figure illustrated, the renormalization factors of both $d_{z^2}$ and $d_{x^2-y^2}$ orbitals decrease monotonically with  double counting values, with the $d_{z^2}$ orbital exhibiting significantly higher sensitivity to double counting adjustments compared to the $d_{x^2-y^2}$ orbital. Experimental measurements of the $d_{z^2}$ orbital's renormalization factor yield a value of approximately 5, corresponding to $E_{dc}^z \sim 10.5$eV in the plotted data. The inset in Fig. ~\ref{fig:density}(b) shows the quantity $\epsilon_{pz}-\mu_{pz}-\epsilon_{loc}^z$ as a function of $E_{dc}^z$, which is proportional to the charge-transfer energy $E_{dp}$. From a theoretical perspective, a lower $E_{dp}$ value enhances interlayer charge transfer phenomena, which manifests computationally as an amplified interlayer term in the Matsubara self-energy. 

\begin{figure}
	\centering
	\includegraphics[width=1\linewidth]{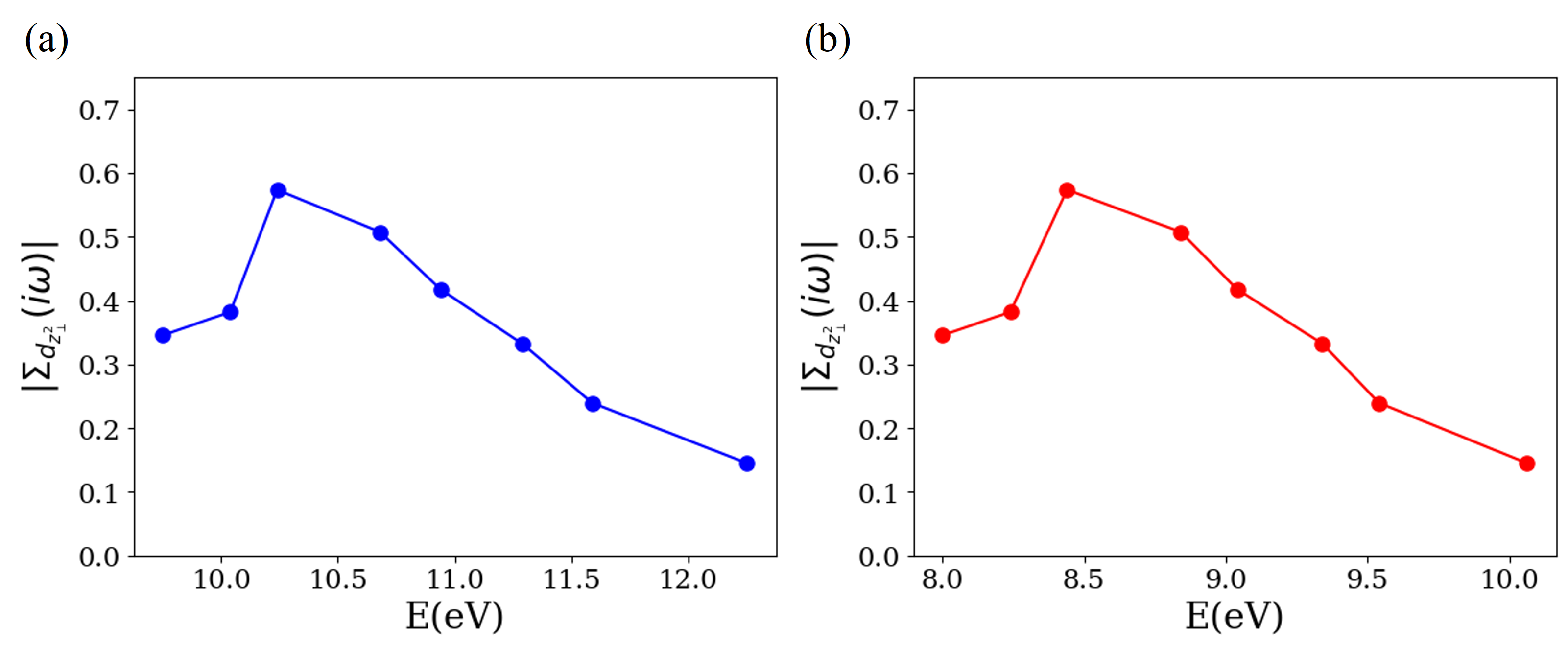}
	\caption{(a),(b) show the interlayer term in $d_{z^2}$ of Matsubara self-energy varying with double counting values in ($E_{dc}^z,E_{dc}^x$). Data points sharing identical self-energy values in panels (a) and (b) correspond to the same parameter set.}
	\label{fig:sigma}
\end{figure}

Fig.~\ref{fig:sigma} investigates the interplay between $E_{dc}$ and interlayer charge transfer dynamics by plotting the Matsubara self-energy interlayer term as a function of ($E_{dc}^z,E_{dc}^x$). Notably, data points sharing identical self-energy values in panels (a) and (b) correspond to the same parameter set. The analysis reveals a similar evolution of the self-energy with respect to both double-counting corrections of  $d_{z^2}$ and $d_{x^2-y^2}$ orbitals. Specifically, in Fig.~\ref{fig:sigma}(a)(Fig.~\ref{fig:sigma}(b)) the interlayer self-energy exhibits a sharp increase at $E_{dc}^z \sim $10.1eV($E_{dc}^x \sim $8.3eV), peaks around $E_{dc}^z \sim $10.2eV($E_{dc}^x \sim $8.4eV), and subsequently decays with further elevation of $E_{dc}$. This non-monotonic behavior deviates significantly from the trend observed in the inset of Fig.~\ref{fig:density}(b). To reconcile this discrepancy, we analyze DOS evolution in Fig.~\ref{fig:dos}(f)-(h). The emergence of metallic states in Fig.~\ref{fig:dos}(g)-(h) disrupts the prerequisite for interlayer charge transfer mediated by oxygen bridges. This breakdown in the charge-transfer pathway explains the divergence between the $\epsilon_{pz}-\mu_{pz}-\epsilon_{loc}^z$ vs $E_{dc}^z$ curve and the self-energy interlayer term's dependence on $E_{dc}$.

To validate against computational results, we evaluated two widely-used double counting estimation methods:(i) FLL form introduced in Ref.\cite{edc_3} , which has the simple form $E_{dc} = U(n_0-\frac{1}{2})-\frac{J_H}{2}(n_0-1) = -14.8$, and $n_0 = 2.44$ stands for the occupancy in $eg$ orbital in no-interacting limit, (ii) Held formulas introduced in Ref.\cite{AMF} with the form $E_{dc} = \frac{1}{3}(U+U-2J_H+U-3J_H)(n_0-\frac{1}{2})=-12.29$. Due to the neglection of energy-level splitting caused by apical oxygen in $e_g$ orbitals, none of these approximations provided distinct double counting values for the $d_{z^2}$ and $d_{x^2-y^2}$ orbitals. The Fermi surfaces calculated by the aforementioned methodology exhibit significant deviations from experimental measurements due to this inherent limitation.

\section{Conclusion}

This study employs CDMFT to investigate the double-counting correction effects on the correlated electronic structure of La$_3$Ni$_2$O$_7$ under ambient pressure, with fixed Fermi surface and adjusted parameters $E_{dc}^z$, $E_{dc}^x$, $\mu_{pz}$ at $\beta =25$. Our calculations reveal a distinct orbital-selective DOS change: the $d_{z^2}$ orbital undergoes the significant variation of DOS near the Fermi level with increasing $E_{dc}^z$, while the $d_{x^2-y^2}$ orbital remains essentially unchanged across the entire $E_{dc}^z$ range. Orbital-resolved electron density calculations reveal different dependence of $d_{z^2}$ and $d_{x^2-y^2}$ orbitals on double-counting corrections, with the $d_{z^2}$ orbital exhibiting significantly stronger sensitivity to double-counting adjustments compared to its $d_{x^2-y^2}$ counterpart. Analysis of renormalization factors shows the monotonic dependence on double counting in both $d_{z^2}$ and  $d_{x^2-y^2}$ orbitals, and it also identifies an optimal $E_{dc}^z$ window (10.2-10.5eV) where $Z_{d_{z^2}}^{-1}= 5-7$ aligns with experimental values, whereas the overestimated $Z_{d_{x^2-y^2}}^{-1}= 3$(vs. experimental $Z^{-1} \sim 2$) likely stems from the Hubbard U=8eV in our model. The analysis of Matsubara self energy reveals a similar evolution with respect to both $d_{z^2}$- and $d_{x^2-y^2}$-orbital double-counting corrections, and it has a nonlinear deviation from theoretical predictions, which we attribute to the metallization of oxygen-bridged pathways that decouples charge transfer efficiency from $E_{dp}$ magnitude. These findings highlight the critical role of double counting parameter selection in modeling nickelate superconductors and provide a computational framework for future studies of layered correlated materials.

\section{Data availability}

The data are available via contacting the corresponding author upon request

\section{Acknowledgements}

See funding support.

\section{Competing interest}

No potential conflict of interest was reported by the authors.

\section{Funding}

This project was supported by  NSFC-92165204, NSFC-12494591, NSFC-92565303, NKRDPC-2022YFA1402802, Guangdong Provincial Key Laboratory of Magnetoelectric Physics and Devices (2022B1212010008), Research Center for Magnetoelectric Physics of Guangdong Province (2024B0303390001), Guangdong Provincial Quantum Science Strategic Initiative (GDZX2401010), and National Supercomputer Center in Guangzhou.

\section{Authors' contributions}

\subsection{Authors and Affiliations}

Guangdong Provincial Key Laboratory of Magnetoelectric Physics and Devices, State Key Laboratory of Optoelectronic Materials and Technologies, Center for Neutron Science and Technology, School of Physics, Sun Yat-sen University, Guangzhou, Guangdong 510275, China\\
Zhong-Yi Xie, Zhihui Luo, W\'{e}i W\'{u}, and Dao-Xin Yao

\subsection{Contributions}

Dao-Xin Yao and W\'{e}i W\'{u} conceived and designed the project. Zhong-Yi Xie wrote the code. Zhong-Yi Xie and Zhihui Luo performed the theoretical calculations and corresponding analysis under the supervision of Dao-Xin Yao and W\'{e}i W\'{u}. Dao-Xin Yao supervised the project and proposed the relevant physical picture. All authors contributed to the interpretation of the results and wrote the paper.

\subsection{Corresponding author}

Correspondence to Dao-Xin Yao





\end{document}